# Superconductivity at ~ 0.049 K in Laves phase HfZn$_2$ predicted by first principles

Md. Zahidur Rahaman[1], Md. Atikur Rahman[2]

[1, 2] Department of Physics, Pabna University of Science and Technology, Pabna-6600, Bangladesh

## Abstract

We report the superconductivity at $T_c$ ~ 0.049 K in Laves phase HfZn$_2$. Based on the first principles method we have studied the details structural and electronic properties of HfZn$_2$ according to which there are 4.67 states eV$^{-1}$ fu$^{-1}$ at Fermi level. Using this value we calculate the specific heat coefficient $\gamma$ theoretically as 10.97 mJ/ K$^2$ mol and then systematically calculate the electron-phonon coupling constant $\lambda$ as 0.45. Finally applying these values in McMillan formula we get the superconducting critical temperature of HfZn$_2$ as approximately 0.049 K.

## 1. Introduction

The Laves phase compounds are considered as promising candidates for superconductivity with widely varying transition temperatures [1]. The study of this compound is so much attractive due to their large differences in transition temperature in the same crystal classes and simple crystal structure. It is still a challenge of condensed matter physics to understand the superconductivity properly in different materials. The chemical composition and crystal structure of Laves phase compound are quite simple and hence the study on these compounds can provide deep knowledge about the nature of superconductivity which motivated us to carry out this investigation. The cubic Laves phase ZrZn$_2$ and HfZn$_2$ have focused a great attention in the recent year. The superconducting critical temperature of ZrZn$_2$ is approximately 1 k [2] which is quite low. Though the details superconducting properties of ZrZn$_2$ have been investigated extensively, the information on HfZn$_2$ is quite missing. The details elastic and thermodynamic properties of HfZn$_2$ have been investigated by Sun et al in 2015 [3] and the details magnetic properties were studied by H.B.Radousky in 1983 [4]. In this work we report the superconductivity in HfZn$_2$ with much lower transition temperature than ZrZn$_2$ by using the plane-wave pseudopotential approach. The remaining parts of this paper are organized as follows. The theoretical methods are discussed in section 2, the investigated results and the related discussions are presented in section 3 and finally, a summary of this present work is shown in section 4.

## 2. Computational Method

We have performed our investigation by using DFT based CASTEP computer program. We have used generalized gradient approximation (GGA) together with the PBE exchange correlation function [5-9]. The pseudo atomic calculation is implemented for Hf-5d$^2$6s$^2$ and Zn-3d$^{10}$4s$^2$. The Monkhorst-Pack scheme is used to construct the K-point sampling of the Brillouin zone [10]. We have used 8×8×8 grids in primitive cells of HfZn$_2$ with the energy cut-off of 400 eV. Geometry optimization has been performed by using the Broyden-Fletcher-Goldfarb-Shanno (BFGS) minimization scheme [11]

..................................................................................................
[2]Corresponding Author: atik0707phy@gmail.com

to obtain the equilibrium crystal structure. The convergence parameters were set to 0.05 GPa for stress, 0.03 eV/ Å for force, $1 \times 10^{-3}$ Å for ionic displacement and $1.0 \times 10^{-5}$ eV/atom for energy.

## 3. Results and discussion

### 3.1. Structural properties

$HfZn_2$ possesses the cubic crystal structure (C15 Laves phases) of space group FD-3M (227). In an unit cell of $HfZn_2$, Zn atoms sit at 16d (0.625, 0.625, 0.625) and Hf atoms occupy the 8a (0, 0, 0) Wyckoff position. The equilibrium lattice parameter of $HfZn_2$ is 7.32 Å [12]. There exist eight formula units in a conventional unit cell of $HfZn_2$. The primitive cell contains two formula units. The lattice parameter and atomic positions are minimized as a function of normal pressure as shown in Fig. 1. The evaluated values of the structural properties of $HfZn_2$ are tabulated in Table 1along with other theoretical and experimental values. It is evident from Table 1 that our calculated lattice constant shows only 0.31 % deviation from the experimental value and exactly same with the other theoretical value bearing the reliability and accuracy of our present calculation.

**Table 1.** Lattice constant "a", unit cell volume "V" and bulk modulus "B" of $HfZn_2$

| Properties | Expt.[12] | Others Calculation[13] | Present Calculation | Deviation from Expt. (%) |
|---|---|---|---|---|
| **a (Å)** | 7.32 | 7.343 | 7.343 | 0.31 |
| **V (Å³)** | - | 395.93 | 395.93 | - |
| **B (GPa)** | - | 115.2 | 170.93 | - |

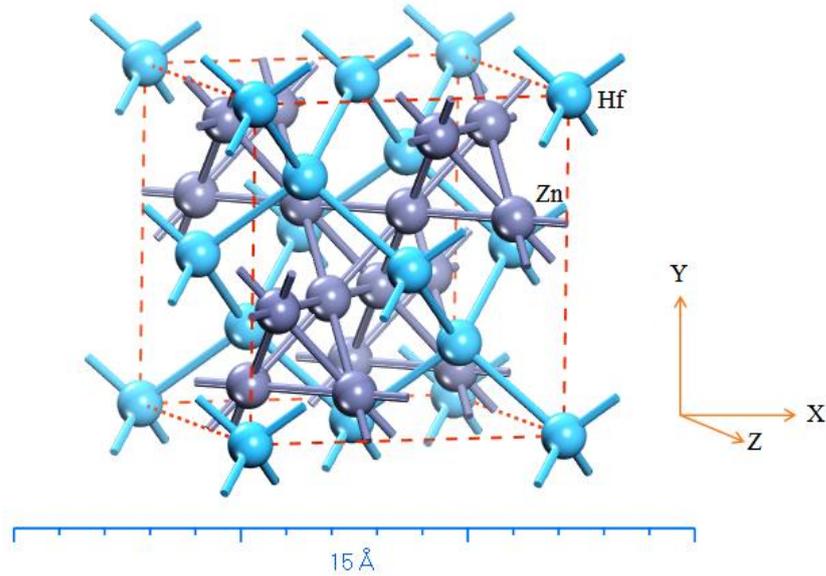

**Fig. 1.** The crystal structures of $HfZn_2$.

### 3.2. Electronic properties

We calculated the band structure, partial density of states (PDOS) and total density of states (TDOS) of $HfZn_2$ as shown in Fig. 2. The main motive for investigating the electronic properties is to evaluate the value of the density of states at Fermi level (N ($E_F$)) of $HfZn_2$ Laves phase. According to the band structure diagram (Fig. 2(a)) the compound under study is metallic in nature since a number of valence bands and conduction bands are overlapped at Fermi level. The metallic nature of HfZn$_2$ indicates that this compound might be a superconductor [14].

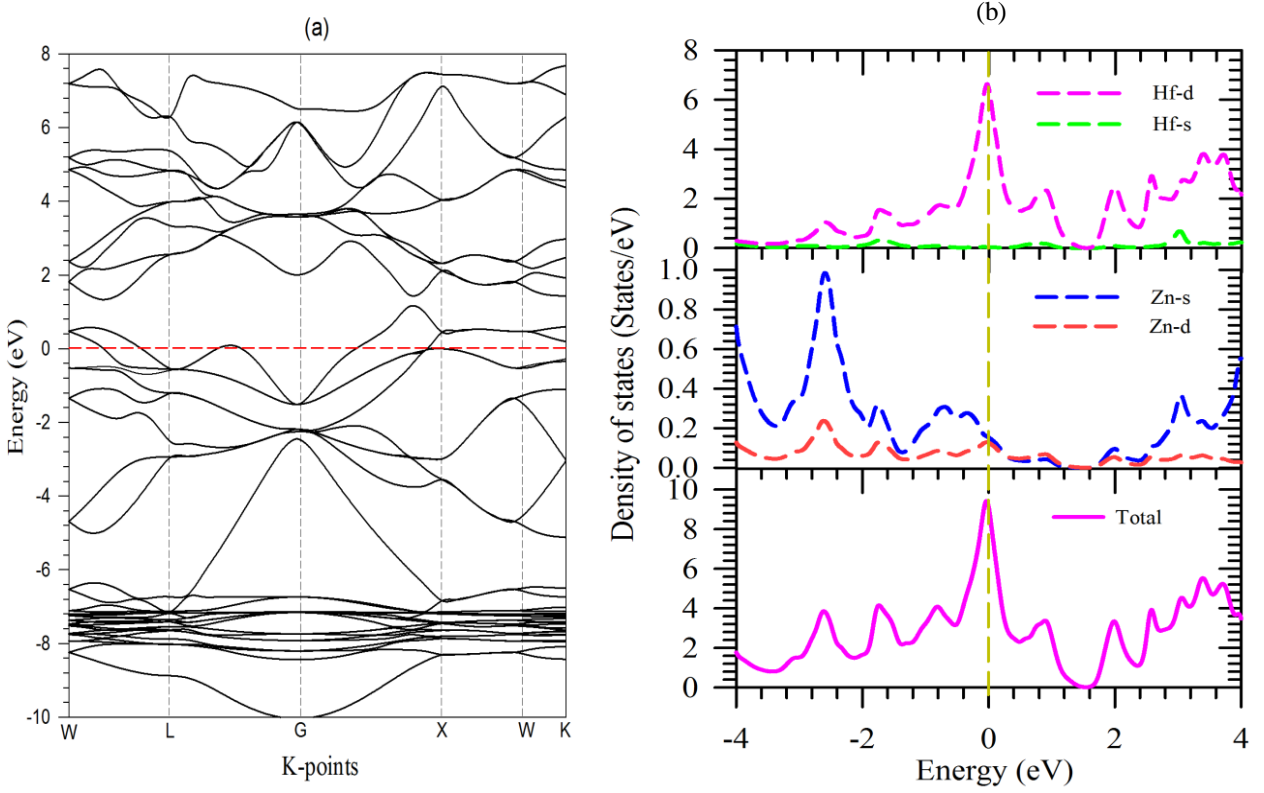

**Fig. 2.** (a) Electronic band structure and (b) Density of states of HfZn$_2$.

According to the DOS diagram illustrated in Fig. 2(b), the main contribution near Fermi level comes from the Hf-d states though both Hf-d and Zn-s band contribute to constitute DOS at Fermi level. From the DOS diagram we calculate the density of state of HfZn$_2$ at Fermi level as 9.34 states/eV cell and 4.67 states eV$^{-1}$ fu$^{-1}$. Our calculated DOS at Fermi level agrees well with the value 4.65 states eV$^{-1}$ fu$^{-1}$ [12].

*3.3. Electron-phonon coupling constant and superconductivity*

Finally we have calculated the superconducting transition temperature of HfZn$_2$ theoretically by using the McMillan formula [15] given by,

$$T_c = \frac{\theta_D}{1.45} e^{-\frac{1.04(1+\lambda)}{\lambda - \mu^*(1+0.62\lambda)}} \qquad (1)$$

Where, $T_c$ is the superconducting critical temperature, $\theta_D$ is the Debye temperature, $\lambda$ is the electron-phonon coupling constant and $\mu^*$ is the coulomb pseudo potential.

The electron-phonon coupling constant $\lambda$ can be obtained by the following equation [16],

$$\lambda = \frac{\gamma_{exp}}{\gamma_{cal}} - 1 \qquad (2)$$

Where, $\gamma_{exp}$ and $\gamma_{cal}$ is the experimental and theoretical values of electronic specific heat coefficient.

The theoretical value of the electronic specific heat coefficient can be calculated by using the following equation [16],

$$\gamma_{cal} = \frac{\pi^2 K_B^2 N(E_F)}{3} \quad (3)$$

Where, $N(E_F)$ is the density of states at Fermi level and $K_B$ is the Boltzmann constant.

Using our calculated value of the density of states at Fermi level in eq.3 we determine the value of $\gamma_{cal}$ as 10.97 mJ/ K$^2$ mole. The experimental value of electronic specific heat coefficient $\gamma_{exp}$ for HfZn$_2$ is 16 mJ/ K$^2$ mole [17]. Substituting these two values of $\gamma_{cal}$ and $\gamma_{exp}$ in eq.2 we have the value of electron-phonon coupling constant $\lambda$ as 0.45 which implies that HfZn$_2$ is a weakly coupled BCS superconductor. The McMillan equation has been found to be accurate for materials with $\lambda < 1.5$ [14]. Coulomb pseudo potential $\mu^*$ can be determined by using the following equation [18],

$$\mu^* = 0.26 \frac{N(E_F)}{1 + N(E_F)} \quad (4)$$

Where, $N(E_F)$ is the density of states at Fermi level. Using our calculated DOS in eq.4 we determine the value of $\mu^*$ as 0.21. Finally using our calculated electron-phonon coupling constant $\lambda$, coulomb pseudo potential $\mu^*$ and Debye temperature $\theta_D = 290$ K [19] in eq. 1 we determine the superconducting transition temperature $T_c = 0.049$ K. All these parameters are shown in Table 2.

**Table 2.** Evaluated specific heat coefficient $\gamma$ (mJ/ K$^2$ mol), electron-phonon coupling constant $\lambda$, the coulomb pseudo potential $\mu^*$ and transition temperature $T_c$ (K) of HfZn$_2$.

|  | $\gamma$ | $\lambda$ | $\mu^*$ | $T_c$ |
|---|---|---|---|---|
| **Cal.** | 10.97 | 0.45 | 0.21 | 0.049 |
| **Exp. [20]** | 16 | - | - | - |
| **Other Cal.[12]** | 10.88 | - | - | - |

## 4. Conclusions

In this work we have predicted a new superconductor HfZn$_2$ of Laves phase family with transition temperature nearly 0.049 K. We have carried out the density functional theory based first principle calculation to determine the density of states at Fermi level and then using this value we have calculated the superconducting critical temperature systematically. We conclude that in this process of determining the transition temperature of the superconductor the value of $T_c$ totally depends upon the value of the density of states at Fermi level. The value of our calculated electron-phonon coupling constant indicates that HfZn$_2$ is a weakly coupled superconductor. Till now there is no literature available on the superconducting properties of HfZn$_2$ and hence it is expected for experimental confirmation of our calculated result soon.